\documentclass[final,3p,times,twocolumn,final]{elsarticle}
\usepackage{graphics}
 \usepackage{graphicx}
 \usepackage{color}
 \usepackage{hyperref}
  \usepackage{morefloats}
\usepackage{amsmath,exscale,verbatim} \usepackage{amssymb} \usepackage[american]{babel}
\biboptions{compress}

\journal{Angewandte Chemie}
\begin{document}
\begin{frontmatter}

\title{Efficient Sonochemistry through Microbubbles Generated with Micromachined Surfaces}

    \author{David Fern\'{a}ndez Rivas*\fnref{labelMCS}}
    \cortext[cor1]{Corresponding author}
    \ead{d.fernandezrivas@utwente.nl}
     
     \author{Andrea Prosperetti\fnref{labelPOF},\fnref{labelP}}
    \author{Aaldert G. Zijlstra \fnref{labelPOF}}
     \author{Detlef Lohse\fnref{labelPOF}}    
    \author{Han J.G.E. Gardeniers\fnref{labelMCS}}

    \address[labelMCS]{Mesoscale Chemical Systems, MESA+ Research Institute, University of Twente. ME147, PO Box 217  7500AE Enschede, The Netherlands.}
%% use optional labels to link authors explicitly to addresses:
 \address[labelPOF]{Physics of Fluids Group, Department of Applied Physics, Faculty of Science, University of Twente. Enschede, The Netherlands}

 \address[labelP]{Department of Mechanical Engineering, The Johns Hopkins University, Baltimore (USA)}

\begin{abstract}
Sonochemical reactors are used in water treatment, the synthesis of fine chemicals, pharmaceutics and others. The low efficiency of sonoreactors have prevented its massive usage at industrial scales. Controlling the appearance of bubbles in place and time is the most limiting factor.
A novel type of sonochemical reactor was designed making use of micro-fabrication techniques to control the nucleation sites of micro-bubbles. The efficiency was increased first by locating the nucleation sites in the most active region of a micro-chamber; additionally the desired chemical effect was significantly higher at the same powers than when not controlled. Silicon substrates were micromachined with ``artificial nucleation sites'' or pits, and placed at the bottom of the micro-chamber. The pits entrap gas which, upon ultrasonic excitation, sheds off a stream of microbubbles. The gas content of the pits is not depleted but is replenished by diffusion and the emission of microbubbles can continue for hours.

Supporting information for this article is available on the WWW
under \href{http://dx.doi.org/10.1002/anie.201005533}{http://dx.doi.org/10.1002/anie.201005533}

Published in: Angewandte Chemie International Edition, 49:50, (2010), 9699-9701
\end{abstract}

\end{frontmatter}

\section{Introduction}
The phenomenon of cavitation, that is, the growth and implosion of gas/vapor 
bubbles in a liquid, is a process which can locally generate extreme 
temperatures of several thousand Kelvin \cite{suslick2008inside} and, for 
this reason, is exploited in sonochemistry to enhance chemical conversion. 
This feature opens the perspective of  high-temperature, high-pressure large
scale systems and therefore holds the promise of constituting a ``green
chemistry'' with a multitude of possible applications in water treatment~\cite{gogate2007application}, material synthesis, the food industry 
\cite{ashokkumar2008modification} and others. 
The ``ideal'' sonochemical reactor from a theoretical point of view is 
a single bubble  trapped in an acoustically driven flask, such as in single
bubble sonoluminescence~\cite{bre02}. There the bubble, in which argon accumulates~\cite{PhysRevLett.78.1359}, collapses periodically
and in a reproducible way, as the collapse does not desintegrate the bubble.
Reactants are sucked into the bubble at expansion and reaction products leave the bubble
at collapse; typical temperatures achieved are 15000~K~\cite{toe03,fla05}.
Another advantage is that thanks to these ideal conditions the problem is accessible to a thorough theoretical treatment whose results are in good agreement with the experimental findings~\cite{bre02,toe03}.
The downside is that the absolute chemical yields are only tiny, as the ambient 
size of such bubbles is in the micrometer regime. For applications, typical sonochemical reactors, such as ultrasonic baths or vessels with 
ultrasound horns attached to their walls, are considerably larger than 
the active region in which cavitation occurs, which is defined by the ultrasound 
field that the transducer generates in the reactor. The difficulty of matching the acoustic cavitation activity with the reactor dimensions so that the complete volume of reactant can be stimulated adds to the low efficiency of most sonochemical processes. Attempts aimed at configuration optimization have mostly had limited success~\cite{yasuda2007enhancement,xu2009dynamical,koda2004sonochemical,tuziuti2006enhancement}.

A new approach is presented to address this problem based on the premise that a significant gain in efficiency may be obtained if the location of the nucleation of bubbles, which subsequently will cavitate due to ultrasound, can be accurately controlled. Achieving this objective will give us control over the spatial distribution of cavitation events, and therewith also over the actual volume of liquid that can be exposed to sonochemical effects. 

The nucleation and formation of bubbles due to ultrasound is mostly heterogeneous, that is, it relies on pockets of gas trapped and stabilized inside randomly existing crevices in container walls or particles suspended in the bulk fluid rendering it extremely difficult to control and predict. The theory developed for bubble nucleation from crevices \cite{atchley1989crevice,borkent2009nucleation} leads the way to control nucleation. As shown in recent work~\cite{bre06,bremond2006interaction,marmottant2006microfluidics,borkent2009nucleation}, stable and monodisperse cavitation nuclei can be formed by trapping gas in pits micromachined in a silicon surface. We use such pits which serve as artificial crevices for the inception of cavitation to achieve higher sonochemical yields at ultrasound powers which would otherwise not produce a significant chemical effect. 

\section{Experimental methods and materials} 
\subsection{Silicon micromachining:} \label{sect:siumachining}
\begin{figure}[t]
            \centering
                        \includegraphics[width=0.6\columnwidth]{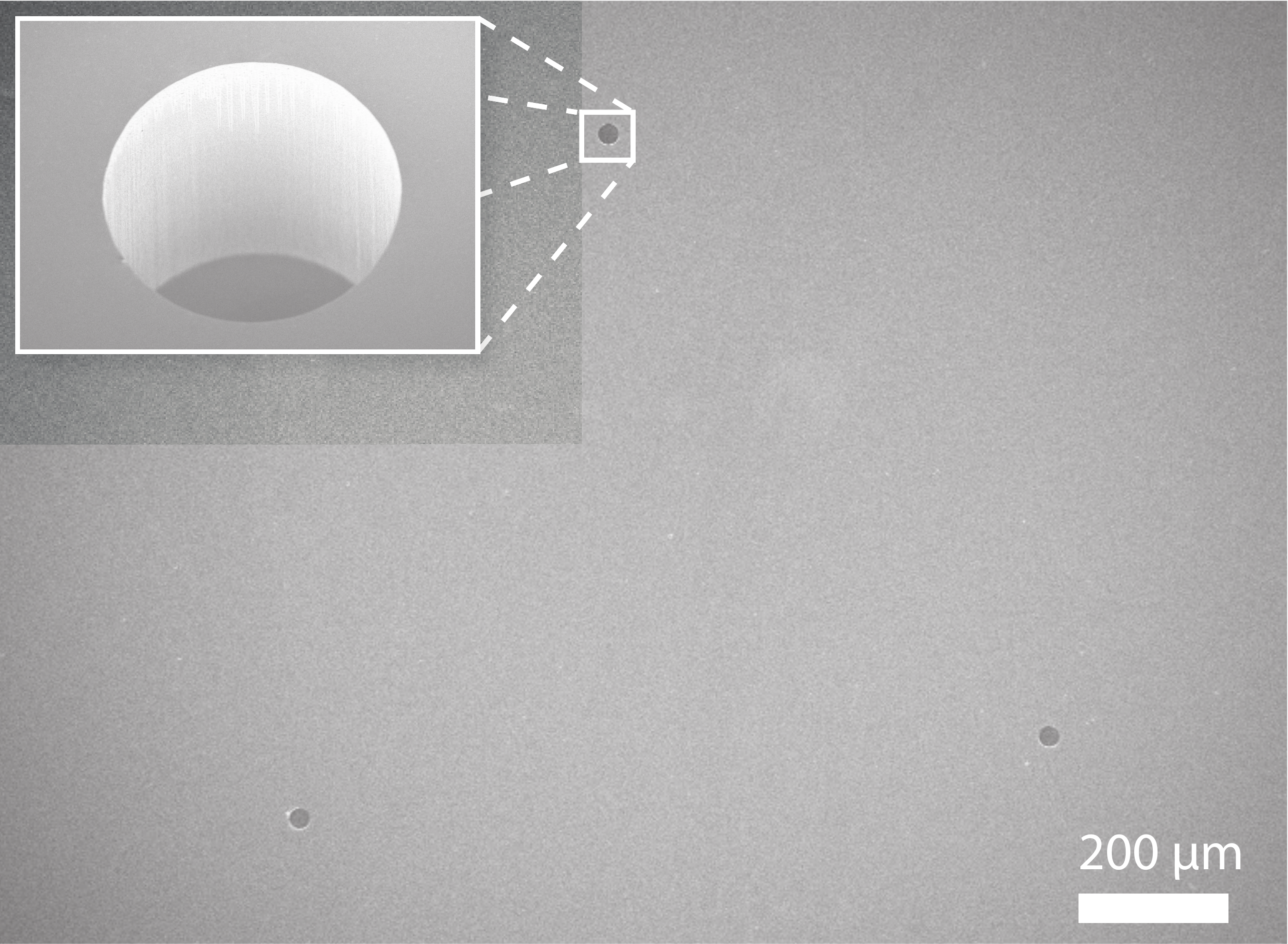}
            \caption{Pits etched in a silicon substrate. Top view of the 3-pit set with 30 $\mu$m diameter micropits.}
            \label{fig:Figure6}
	\end{figure}

Three different configurations of pits were used. The pits had the same diameter 
(30 $\mu$m) and were arranged in sets of one, two (in a line) or three 
(in a triangle, see Figure \ref{fig:Figure6}) at a distance of 1000 $\mu$m from each other.

The substrates were micromachined under clean room conditions on double-side polished silicon wafers and spin coated with the photosensitive substance Olin 12, on which the designed pattern was transferred with a mask aligner EV620 (photolitography). After development the pit pattern was open and with a plasma dry-etching machine Adixen AMS 100 SE (Alcatel) process BHARS, the holes were etched into the silicon substrate at the desired depth. The machined diced silicon
 square pieces of 1 cm-side were mounted to the bottom of a small glass container which contained a liquid volume of 300 $\mu\ell$, to the bottom of which a piezo element was attached.

\subsection{The acoustic field}
\begin{figure}[t]
    \centering
        \includegraphics[width=0.85\columnwidth]{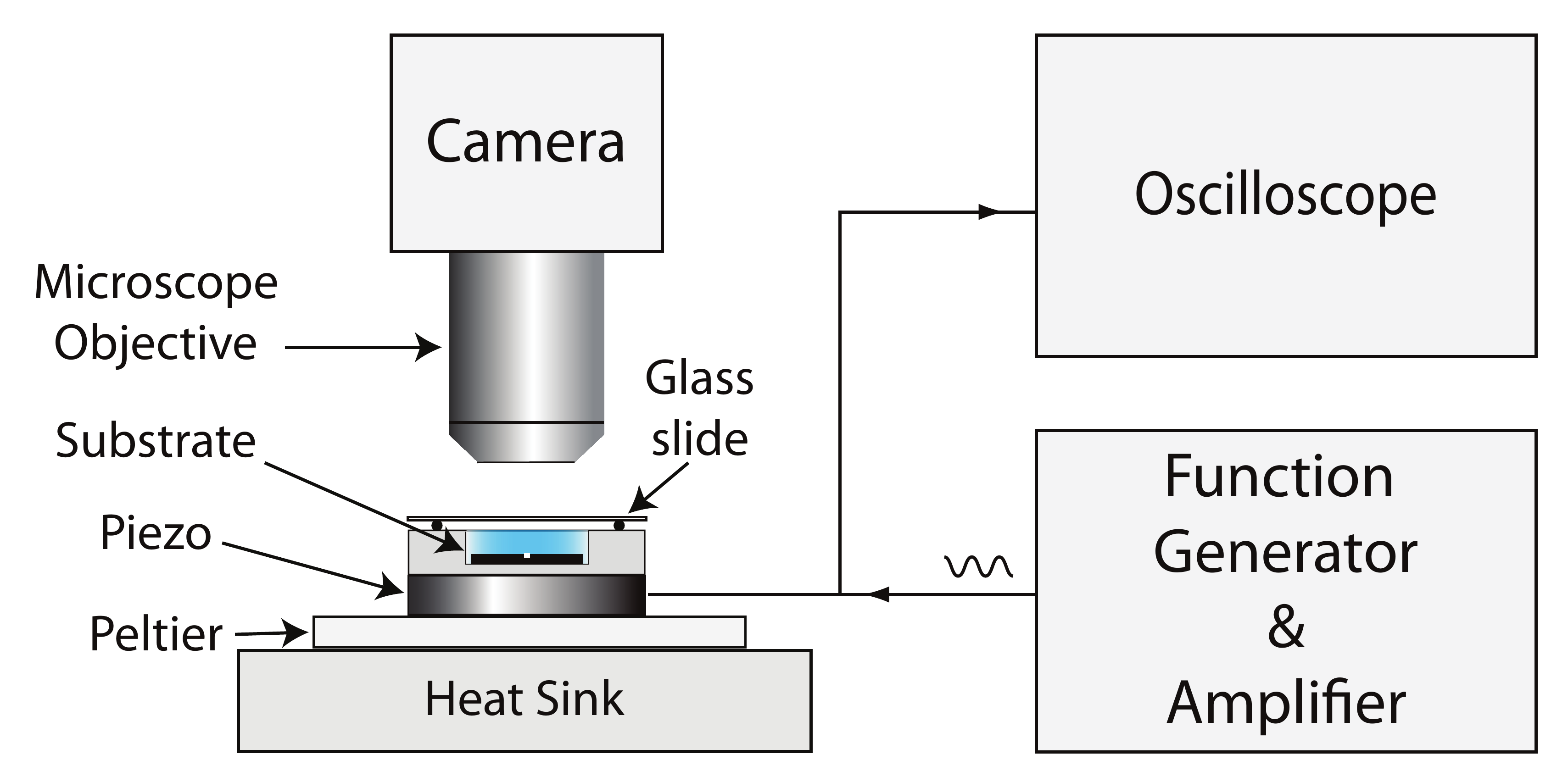}
          \caption{Experimental setup}
          \label{fig:Figure7}
     \end{figure}

The reaction chamber was a glass container of 2.5~cm outer diameter, 1.5~cm 
inner diameter and depth of 3~mm, and bottom thickness of 2~mm. the bottom 
thickness matched the quarter-wavelength vibration imparted by a piezo Ferroperm PZ27 6 mm thick with diameter of 2.5~cm, glued to the bottom of the reaction chamber. The top of the glass container was capped with a rubber ring and a glass slide to avoid evaporation of the sample during the experiments (see Figure \ref{fig:Figure7}). 

%\newpage
A sinusoidal standing acoustic field of 200 kHz was generated by a \emph{Hewlett Packard} model 33120A arbitrary waveform generator, amplified with a \emph{SONY TA-FB740R QS} amplifier. The voltage, current and phase difference provided to the piezoelectric element were measured with a \emph{Tektronix} DPO 4034 oscilloscope and \emph{Tektronix} current probe CTA-2.  

Three different power settings were used: low (74 mW), medium (182 mW) and high (629 mW). During the experiments, the bottom surface of the piezoelectric element was in contact with a Peltier element (Marlow Industrial) to keep the temperature at a constant value of approximately 25$^{o}$C.

\subsection{Luminol visualization and determination of the radical formation rate}
The homolytic cleavage of $\text{H}_{\text{2}}\text{O}$ by sonolysis was investigated using \emph{Luminol} and \emph{terephthalic acid}.  Luminol is a well-established indicator for the visualization of active sonochemical regions, because it reacts with OH$^.$ radicals, whereupon reaction gives luminescence with an intensity proportional to the amount of radicals produced \cite{henglein1989luminescence}. A solution of 1$\times \text{10}^{-\text{3}}$ mol/$\ell$ Luminol and 1$\times \text{10}^{-\text{4}}$ 
mol/$\ell$ hydrogen peroxide  was prepared with adjusted pH=12 by adding 
$\text{Na}_2 \text{CO}_\text{3}$, as described in \cite{felver2009cavitation}. 

Conversion of terephthalic acid to 2-hydroxyterephthalic acid (HTA) was taken 
as a quantitative measure for the concentration of OH$^.$ radicals formed by the ultrasound induced bubble activity. A calibration graph for the fluorescence intensity as a function of HTA concentration was obtained following steps similar to those described in \cite{Mason1994}. Fluorescence was measured using a spectrofluorometer (\emph{TECAN Sapphire}) with an excitation wavelength of 310~nm and an emission wavelength of 429 nm. A graph of fluorescence intensity against HTA concentration was plotted and gave a straight line of positive slope for concentrations. The Terephthalic acid solution to be used as dosimeter was prepared by mixing 0.332 g of Terephthalic acid  (\emph{Sigma-Aldrich}, 2.0$\times \text{10}^{-\text{3}}$ mol/$\ell$), 0.200 g of NaOH (5.0 $\times \text{10}^{-\text{3}}$ mol/$\ell$), and phosphate buffer (pH 7.4), prepared from 0.589~g of $\text{KH}_\text{2} \text{PO}_\text{4}$
(4.4 $\times \text{10}^{-\text{3}}$ mol/$\ell$) and 0.981~g of $\text{Na}_\text{2} \text{HPO}_\text{4}$ ($\text{7.0} \times \text{10}^{-\text{3}}$~mol/$\ell$, all from \emph{Riedel - de H\"aen}). The resulting solution was then made up to 1 $\ell$ with water~\cite{Iida2005}.  For each experiment run,  the same amount of terephthalic acid solution 300 $\mu$$\ell$ measured with \emph{Eppendorf} micropipettes was used. 

At the end of the experimental run, the solution was pipetted out of the reaction chamber and stored in the dark in sterilized vials (manufactured by \emph{Brand}) for ultrapure chemical analysis for further spectroscopic analysis. 
Later, 200 $\mu$$\ell$ taken from the samples were pipetted into the wells of an assay plate (\emph{Corning Inc.}) to be analyzed with the spectrofluorometer described above. The conditions for sample analysis were: gain 40, height from the well, 9000 
$\mu$m, integration time 100 $\mu$s, 10 reads per well. Excitation was at 310 nm 
and the emission scan was from 350 to 600 nm. Each experiment was repeated at least 6 times showing acceptable consistency.

Calculation of the radical formation rate was done according to:
\begin{equation}
            \frac{dN_{rad}}{dt}=\frac{(c_{end}-c_{initial})N_{A} V}{\Delta t}
\label{eq:dndtradicals}
\end{equation} 

Here $c_{end}$ and $c_{initial}$ are the final and initial concentration in moles per volume, $N_{A}$=6.02$\times10^{23}$ mol$^{-1}$ is Avogadro's number and $V$ is the liquid volume exposed to the ultrasound (300 $\mu$ $\ell$). With $c_{initial}=0$ this relation provides an estimate of the rate of radical formation for all pits and power combinations.

\section{Results}
\begin{figure}%[h]
	\centering
		\includegraphics[width=0.8\columnwidth]{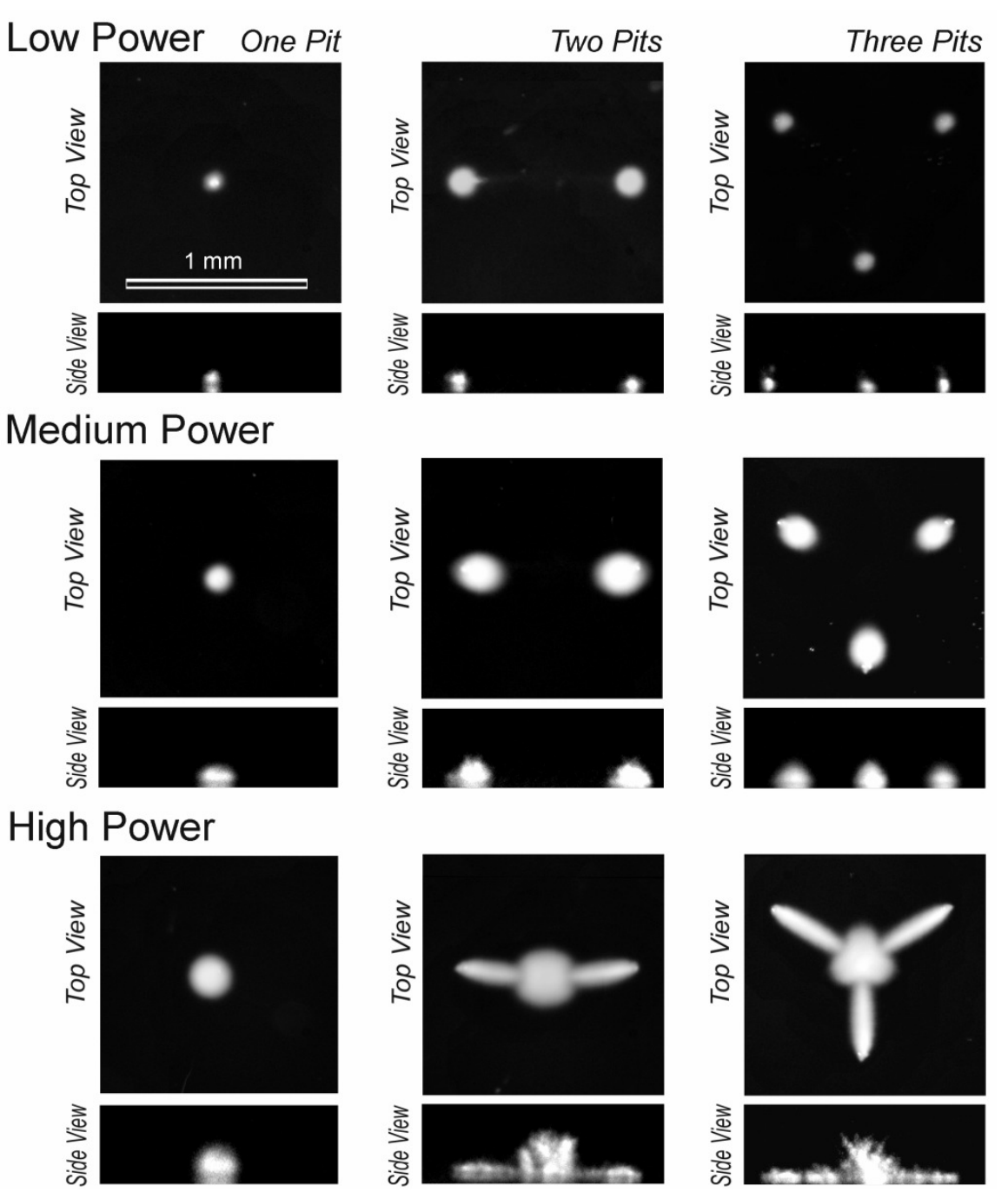}
\caption{Images showing both the top and side view of the bubble structures generated from the micropits for different configurations (1,2 and 3 pits) and for increasing power level. Low power corresponds to 74 mW, medium power  to 182 mW, and  high power to 629 mW.}
	\label{fig:Figure1}
\end{figure}

The micropit bubble ejects streams of microbubbles, which resemble previously reported streamers ~\cite{leighton1995bubble,neppiras1980acoustic}. The observed trajectories in Figure \ref{fig:Figure1} are the result of the complex interplay of primary and secondary \textit{Bjerknes forces} and microstreaming.

For a single pit, at low and medium ultrasound powers, the microbubbles are ejected and oscillate several cycles until they dissolve or are recaptured by the micropit bubble. 
 For low power the situation is similar for the configurations with two and three pits.  With increasing power a drastic transition in the bubble flow pattern is observed. Beyond the transition the microbubbles move away from the symmetry axis 
of their respective pits and towards one another (see intermediate row in Figure~\ref{fig:Figure1}). The microbubbles then form a dense bubble cloud traveling towards the midpoint between the two-pit or to the midpoint of the three-pit arrangement. For the three-pit configuration the microstreamers point to the center of the triangular array and form a triangular cloud of bubbles, as seen in top view. 

\begin{figure}[!htb]
	\centering
		\includegraphics[width=0.8\columnwidth]{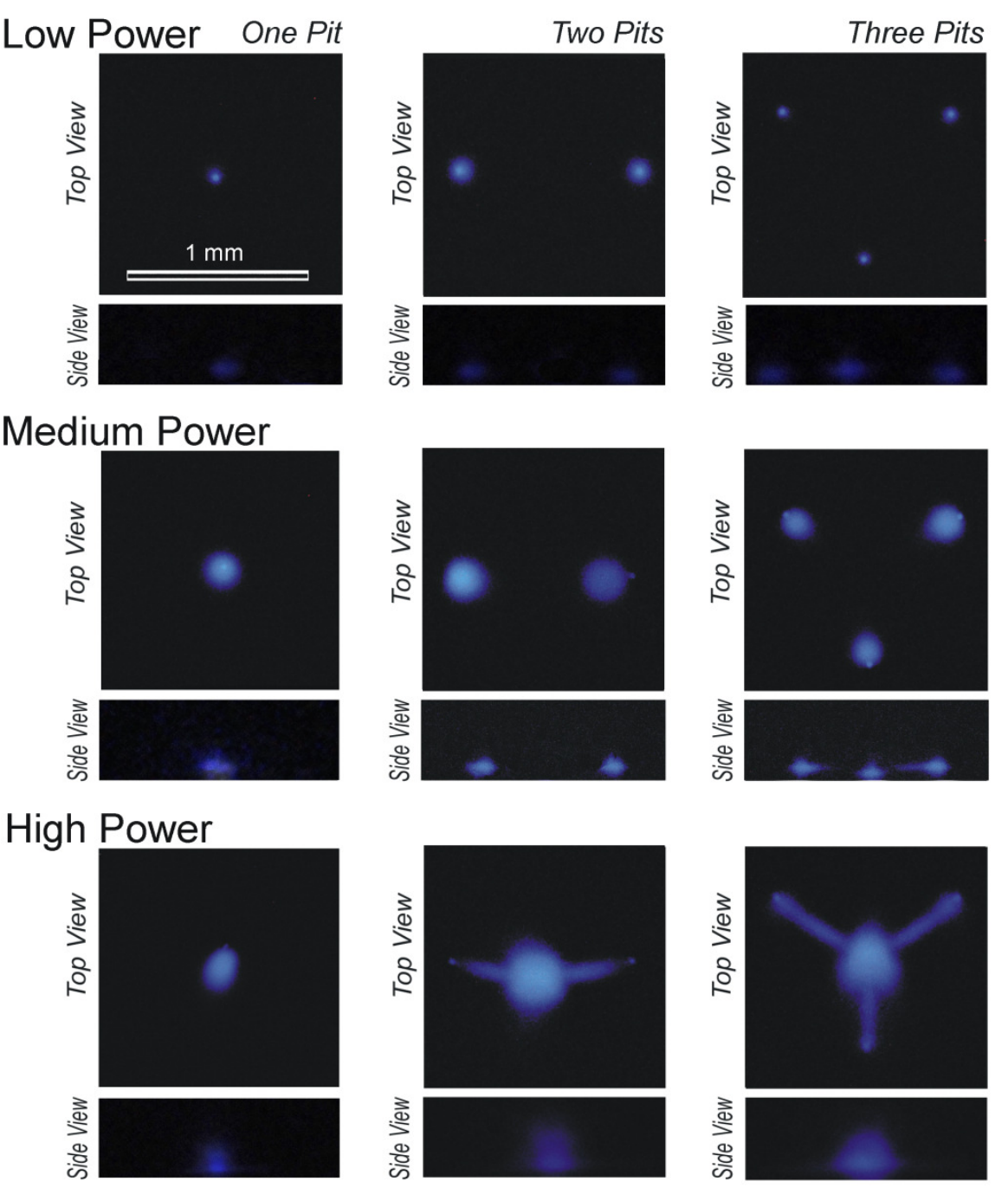}
	\caption{Luminol luminescence in dark room conditions. Images showing both the top and side view of the bubble structures generated from the micropits for different configurations (1,2 and 3 pits) and for increasing power level.  Compare with Figure~\ref{fig:Figure1}}
 \label{fig:Figure2}	
\end{figure}

These experiments were repeated with the water solution of luminol.  In Figure \ref{fig:Figure2} the resulting images can be seen. 

Imaging of the low intensity of the emitted luminescence required the use of long camera exposure times (typically 30 seconds). A detailed comparison of the features in Figure \ref{fig:Figure2} with those in Figure \ref{fig:Figure1} shows a perfect match. Clearly, the regions of cavitation activity correspond to the regions where light is emitted due to sonochemiluminescence~\cite{henglein1989luminescence}.

\begin{figure}[htb]
	\centering
		\includegraphics[width=0.9\columnwidth]{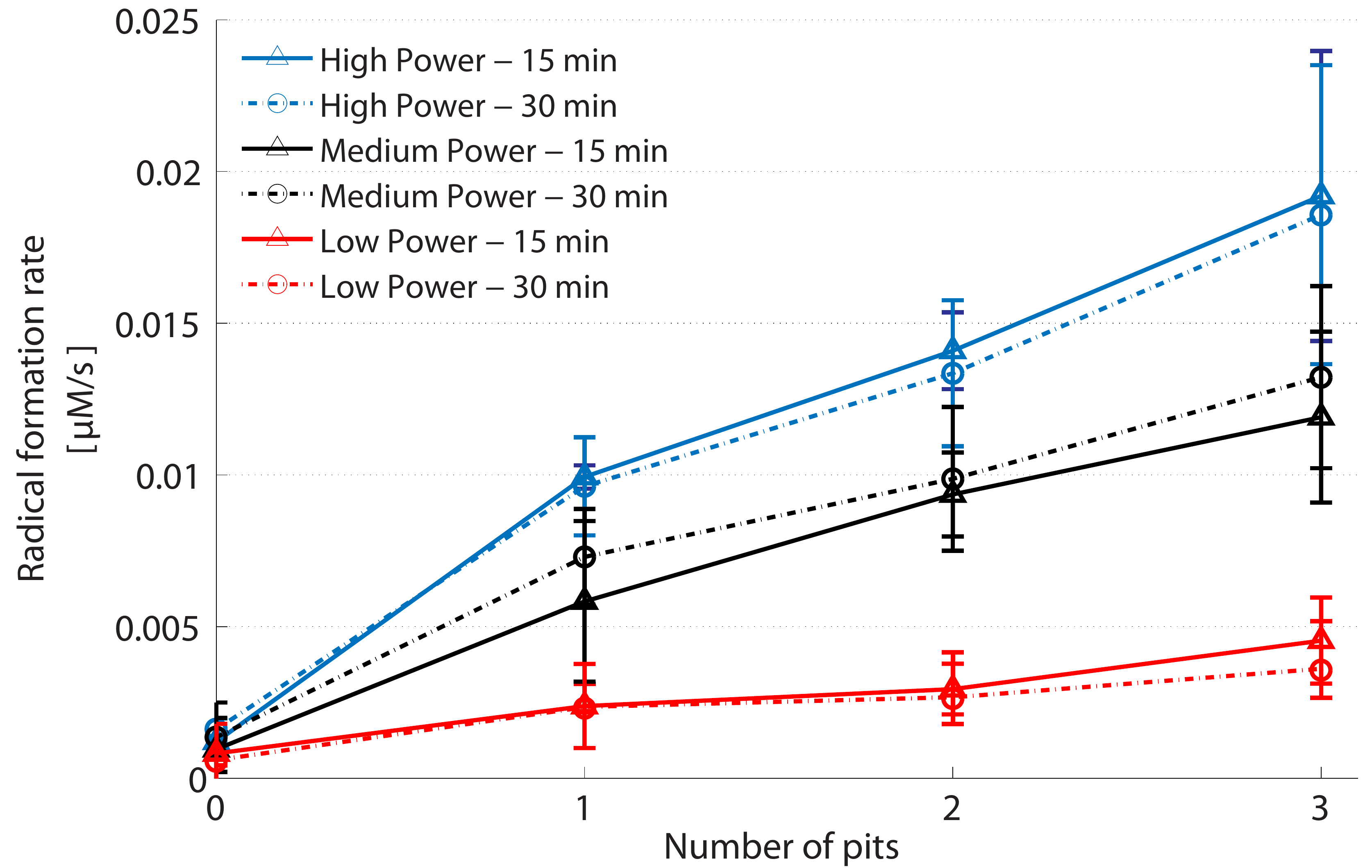}
	\caption{Radical formation rate vs. pits. Low power=74 mW, medium power=182 mW and high power=629 mW. The bars show the range of the experimental data}
	\label{fig:Figure3}
\end{figure}

For a quantitative measure of radical formation, the terephthalic acid dosimetry method was used. The fluorescence intensity emission of HTA at 429 nm (measured ex-situ) allows us to calculate the amount of radicals generated by ultrasound induced microbubble cavitation. This measurement was performed for the three pit configurations, each at low, medium and high power and for 15 and 30 minutes process durations. In addition, experiments with the same conditions were done with a silicon chip without micropits. These measurements showed no significant evidence of radical formation. The radical generation rates (equation \ref{eq:dndtradicals}) resulting from off-line fluorescence measurements are shown in Figure~\ref{fig:Figure3}. 

The data show that the presence of pits gives a significant enhancement of the radical formation rate. The rate is stable over time and is an increasing power of the ultrasound field as expected. Radical formation increases with the number of pits, but the slope of the lines decreases as more pits are added, presumably due to the onset of interactions with the bubble clouds generated by the other pits. 
At high power, the bubble pattern changes dramatically (Figures \ref{fig:Figure1} and \ref{fig:Figure2}) and is expected to lead to a different radical generation distribution over the reactor volume. The quantitative interpretation of these data is complex. Generally speaking, smaller bubbles are stiffer due to a surface tension contribution and may not grow as large during expansion, with a consequent weaker compression and lower maximum temperatures. Large bubbles, on the other hand, may not collapse spherically, especially when close to a solid surface, which would also limit 
their maximum compression. It is not clear, however, to what extent these
general trends are relevant in the experimental situation of present
concern. 

The most important parameter for the evaluation of our results and comparison with the work of others is the energy efficiency defined as:
\begin{equation}
X_{US}=\frac{\Delta H }{P_{US}} \frac{dN_{rad}}{dt}
\label{eq:effieq}
\end{equation}
where $\tfrac{dN_{rad}}{dt}$ is the radical formation rate in moles per second, 
$\Delta H$ is the energy required for the formation of OH$^.$ radicals, 
which is equal to the enthalpy of formation associated with the following 
chemical reaction:
\begin{equation}
H_2O \stackrel{\Delta H= 5.1 eV}\rightleftharpoons OH^{.} + H^{.} 
\label{eq:formation}
\end{equation}
This enthalpy has a value of  5.1 eV per molecule~\cite{toegel2002suppressing}. $P_{US}$ is the electric power absorbed by the transducer which can be obtained from the measured voltage, current 
and their phase difference. 
The efficiency as defined in Equation \ref{eq:effieq} is shown in Table \ref{tab:XUS} for each of the configurations studied. 

		\begin{table}[!htb]
		\centering 
		\samepage
			\caption{Efficiency $\times10^{6}$ as defined in equation \ref{eq:effieq}}
			\begin{tabular} {@{}cp{0.4in}p{0.4in}p{0.4in}p{0.4in}@{}}   
			 \hline	%\toprule		 
			 High power& & & & \\
			 629 mW	& 3 pits  & 2 pits	 & 1 pit    & 0 pit	\\	\hline
		 	 15 min         & 4.5 & 3.3 & 2.3& 0.3	  	\\			%\hline
			 30 min 				& 4.4 & 3.1 & 2.3& 0.4	  	\\			%\hline
			  \hline	%\toprule		
			 Medium power& & & & \\
			 182 mW&3 pits & 2 pits& 1 pit    & 0 pit	\\	\hline
		 	 15 min & 9.7 & 7.6 & 4.7  & 0.8	  	\\			%\hline
			 30 min & 11 & 8.0 & 5.9  & 0.1	  	\\			%\hline
			  \hline	%\toprule		
			 Low power& & & & \\
			 74 mW& 3 pits & 2 pits& 1 pit & 0 pit	\\	\hline
		 	 15 min & 9.1 & 5.9 & 4.8  & 1.6	  	\\			%\hline
			 30 min & 7.1 & 5.2 & 4.6  & 1.1	  	\\			%\hline
	\hline		%		\bottomrule
		\hspace{3 mm}
			\end{tabular} \label{tab:XUS}
	\end{table}

\section{Discussion and Conclusions}	
From the values in Table \ref{tab:XUS} we can draw the following conclusions: (i) Comparing the efficiency values of the chips with one or several pits with those of the chip with no pits, there is an efficiency increase by an order of magnitude, which demonstrates that the pits on the reactor wall give a considerable enhancement. (ii) The efficiency obtained even with such a low number of pits is close to the highest efficiencies reported in the literature with conventional sonochemical reactors \cite{Didenko2002,kuijpers2002calorimetric,rochebrochard2009comparison,sutkar2009design,hallez2009characterization,mark1998oh,Iida2005,mandroyan2009modification2}. A detailed comparison is hardly possible due to the use of different frequencies, chemical dosimeters and other physical parameters. 

From the values of Table \ref{tab:XUS} it can be seen that the most efficient setting is medium power in all pit configurations. Several factors may explain this observation. At high power, the microbubble collapse is more catastrophic which results in a smaller compression and heating of the gas. In the case of two- and three-pit configuration, the microbubbles stay longer in the most active zone close to the substrate while traveling parallel to the surface; but the collapse of microbubbles close to the surface is highly non spherical, again reducing the maximum compression and heating of the gas. 
The microbubbles in the dense cloud at the midpoint may also have a weaker collapse due to the shielding by the outer microbubbles.

\begin{figure}[htbp]
            \centering
                \includegraphics[width=0.9\columnwidth]{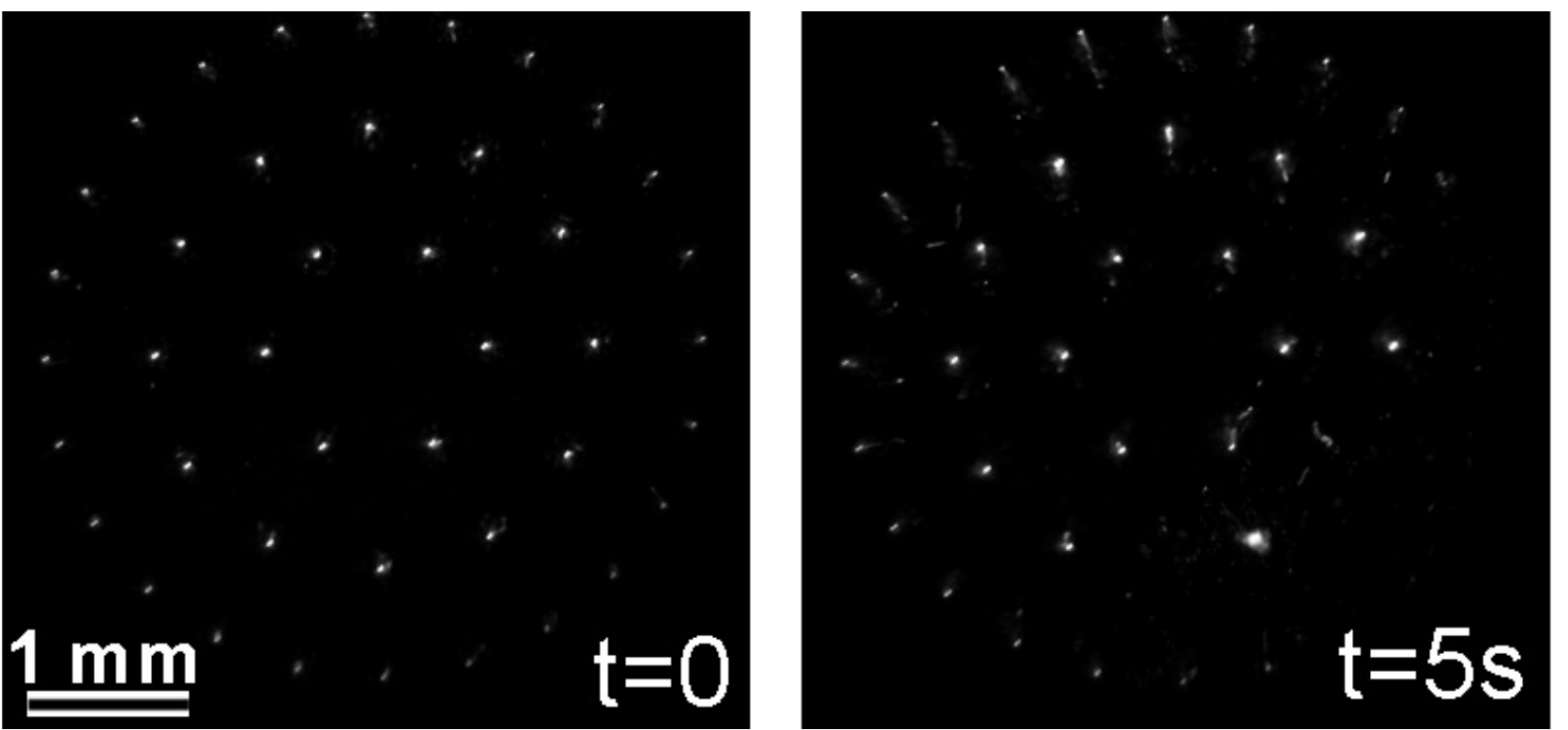}
            \caption{Microscope image under normal light conditions showing an arrangement with a higher number of pits (42). Here similar ejections of microbubbles from the micropit bubbles occur.}
            \label{fig:Figure5}
\end{figure}

It has to be stressed that our results were obtained with only a very 
small active region, with state-of-the-art microfabrication procedures 
it is straightforward to increase the number of pits (see Figure \ref{fig:Figure5}). Based on our results in Table \ref{tab:XUS} and Figure 
\ref{fig:Figure3}, increasing the number of pits should give 
an increase in energy efficiency by a factor of $100$ or more. This estimate is based on a linear extrapolation of the one-, two- and three-pit configuration at medium power settings.

Our results indicate that the introduction of micromachined pits on the 
surface of a reactor wall attached to an ultrasound transducer can boost 
sonochemical efficiency. The underlying mechanism is the production of 
microbubble cavitation clouds from the stable gas pockets that form in these pits.

\section*{Acknowledgements}
This research was supported by the Technology Foundation STW, Applied Science Division of NOW and the Technology Programm of the Ministry of Economic Affairs, The Netherlands.

\section{References}
%% From Bram's
%\bibliographystyle{../jasanum}
%{\scriptsize\bibliography{BiblioAngew}}

% from AAldert's
%\clearpage
%\putbib

\end{document}